# Self-Healing by Means of Runtime Execution Profiling

Mohammad Muztaba Fuad, Debzani Deb, Jinsuk Baek

Dept. of Computer Science, Winston-Salem State University, Winston-Salem, NC 27110, USA

{fuadmo, debd, baekj}@wssu.edu

**Abstract**

*A self-healing application brings itself into a stable state after a failure put the software into an unstable state. For such self-healing software application, finding fix for a previously unseen fault is a grand challenge. Asking the user to provide fixes for every fault is bad for productivity, especially when the users are non-savvy in technical aspect of computing. If failure scenarios come into existence, the user wants the runtime environment to handle those situations autonomically. This paper presents a new technique of finding self-healing actions by matching a fault scenario to already established fault models. By profiling and capturing runtime parameters and execution pathways, stable execution models are established and later are used to match with an unstable execution scenario. Experimentation and results are presented that showed that even with additional overheads; this technique can prove beneficial for autonomically healing faults and reliving system administrators from mundane troubleshooting situations.*

## I. INTRODUCTION

Today's computing environments are complex and heterogeneous tangle of hardware, middleware and software from multiple vendors that is becoming increasingly difficult to program, integrate, install, configure, tune, and maintain. This leads to the idea of autonomic computing [1], [2] where the complexity and the management of such systems is handled by the system itself. One aspects of such self-management is self-healing where the system corrects itself after a failure situation. In real life, the end users of today's big and complex software are left with the task of managing the system when the computational task is faltering due to failures that they cannot fix because of lack of computing knowledge. To provide users with a self-managed system and to achieve the goal of self-healing, we need to provide two supports. Firstly, since we do not want the developers worry about incorporating the self-healing features, appropriate code transform and injection techniques are needed at the object code level (merely because for older systems, the source code may not be available). Secondly, there should be a way to identify a specific failure and distinguish it from other similar failures with different root causes. None of these tasks are trivial and requires extensive effort. Our past experience on code transformations [3], [4] showed that the first of the above mentioned support is achievable. In this paper, we are going to present an approach to correctly identify and distinguish failures in a software application so that self-healing can be provided autonomically. This support is necessary if we want to envision a true self-healing system where the self-management feature will actively categorize new faults, learn patterns for cause-affect relationships, perform management actions for identified faults and even try to predict what to do in case an unknown fault appears in the system.

The rest of the paper is organized as follows. Section II talks about the motivation behind this work and discusses on related work. Section III gives an overview of the system and the approach. Section IV discusses about fault similarity and relationship. Section V shows results from experimentations and finally Section VI concludes the paper.

## II. MOTIVATION AND RELATED WORK

The motivation of this research comes from the fact that it is a non-trivial task to automate the process of making a regular user application into a self-healing application. To achieve this goal, proper code transformation techniques (to add required functionalities) are needed. Our previous investigation [3], [4] in this research area showed that it is viable to do code transformations to inject autonomic properties into existing applications. It also showed that relieving the end-user from the complex programming interfaces and metaphors, the application is more eligible to users who do not have advanced knowledge in programming paradigms.

When a runtime software system requires management due to any kind of failure or user action, an effective fix (management action) should be identified and applied quickly. Failure can occur due to system faults or due to performance bottlenecks. In most cases, the root cause of such failure is human ignorance or mismanagement. It is therefore compelling to have systems that self-manage itself and reduce the day-to-day involvement of humans in the operation of the system. Matching a fault with the database of existing faults is also not trivial [5] since the same fault can occur for multiple different reasons and in different circumstances. By keeping track of what the program is doing during fault free runs and comparing that with situation where the program is not running smoothly; fault models can be generated. Also, after a few times of interactions with a particular software application, users naturally express a great deal of their goals, preferences, and personality. That way, patterns of usage and failures scenarios can be learned over time by monitoring such application.

In the last couple of years, software fault healing (specially, adaptive or self-*) was forced back to the spotlight because of current software's inherent complexity and requiring more expert human intervention and man-hour to manage and maintain these software. There is a plethora of research work addressing this issue and some of the research related to this work is presented here. One major difference between the work presented in this paper and other related work is that we establish domain specific inherent relationships among faults in a system to find the root cause of a fault by looking in to the similarity/dissimilarity between faults by using an established execution trace history of the program. We believe that, building up these relationships between faults and root causes will allow the system us to match fault scenarios faster and allow systems to provide true self-healing features.

Ding, et al. [6] proposes a black-box approach of software development that automatically diagnoses several classes of application faults using the application's runtime behaviors. Their approach collects application's runtime signature as we do, but instead of concatenating previous execution traces to form signatures, we generalize traces and formulate signature. Also, instead of manual invocation of diagnosis process, we incorporate that as part of the system (by means of code injection), so that in an unstable state, the injected codes try to diagnose and solve the problem itself with minimum or no human intervention.

Yuan, et al. [7] try to find correlation of an unstable application state with a list of solved cases. Instead of using vague text descriptions to identify problem situations, the authors employ statistical techniques to match an unknown fault to a set of known fault situations. An obvious difference with our approach is that we trace the target application in a regular execution to establish known execution paths and signatures and once an unknown case is identified the injected code handles all related healing procedures. In this work, an already establish list of 100 top faults and their root causes are given (specific to a certain problem domain) and an unknown cases is matched with this list of cases and the one which is close to the unknown fault is provided to the user so that the user can take control and solve that particular situation. Cook, et al. [5] presents a similar approach that directly matches an unknown fault to a list of known fault. We think both of these work are good foundation for further extension and we utilizes both these ideas in our approach and extended by creating relationship between faults and signatures and also by automating the process of healing the software once the fault is diagnosed.

Binkley, et al. [8] presented a work where the authors used text similarity measures to predict fault from existing application log data. Although their approach is innovative, we cannot employ that in our problem domain because of lack of fault data and logs and because of dissimilarity between the underlying data.

## III. SYSTEM OVERVIEW

Fig. 1 shows a general overview of the application life cycle. The user application is statically analyzed and appropriate code segments and hooks are injected on the object code and a modified (the computation logic of the user application remains same) version of the application is produced, which acts as an independent entity (autonomous entity). Although the figure shows the code inserts and the original code of the same size, there is no relationship and the overall code inflation depends on different factors. More on code transformation techniques and code inflation was discussed in [3], [4], and [9].

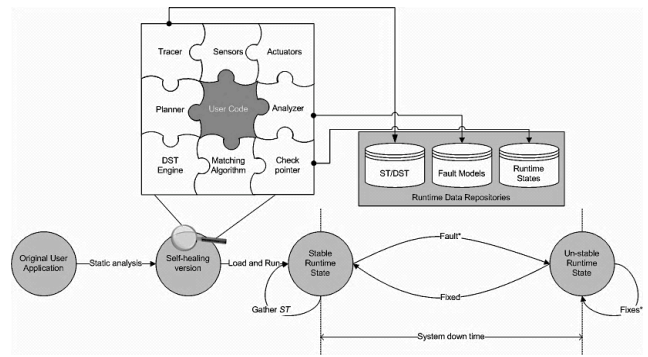

Fig. 1 Application's life cycle.

Major characteristic of the code transformation process is as follows:

- Java byte code is used as the target for application's programming domain.
- The code transformation uses standard OOP metaphors (Class, Object and Methods) to inject code segments. Therefore, with minor modification, the techniques can be ported to other interpreted OOP language.
- The granularity for adding self-healing is per method basis.

Once the static analyzer produces the transformed code, it is executed in the similar fashion as the original program. The modified program runs in two different states:

a) *Stable State*: In stable state, the tracer keeps all records of execution and establishes signature of stable execution paths of the program. The check-pointer keeps track of the application in regards of its current states (field values, object states, execution point etc.), so that once the software needs to restore to the last stable state, it can do that after healing whatever failure it encounters. The modified application collects runtime parameters (signature) and execution pathways (traces) and stores it in an object called *Signature-Trace* or *ST*. A *signature* of a program is its runtime state including field values, object states, open files, environment information etc. We can represent this information as an *n-gram* based representation as in [6]; however will not be adequate to hold the relationships and rank of those attributes for generalization purpose. A *trace* is program execution path that records method invocation and stack traces. After a stable run, the acquired signature and trace is merged into a data structure that we are denoting as *Signature-Trace* or

*ST*. After each stable run, the generated *ST*s are compared and generalizations are made. To make generalizations, we do not just pick the common entities or generalize on the value of a field or parameter. We rank each entity by their occurrences and sort them accordingly. However, to limit the inflation of the data structure, a threshold is set and if the size goes beyond the threshold limit, the least ranking entity is deleted. *ST*s are created for every single run of the application and in regular intervals are shared among the machines in the system. Each machine merges all gathered *ST*s and generates *Distributed ST* or *DST*. Initially, the system treats *DST*s as the sole source for domain knowledge. However, with enough run of the application and generation of enough *DST*s, generalizations are made and global *Domain Knowledge* (*DK*) database is generated from which fixes have to be deduced for different fault scenarios. It is to be noted that, once *DK* is generated, each machines need to update its *DST* with the more generalized version of any *ST* combinations from *DK*. Eventually, well-formed fault scenarios are generalized and put into the fault model database. This whole process is shown in Fig. 2. One point to be noted from Fig. 2(a) is that the target application has to have different running instances for faster generations of *DST*s. Also, in Fig. 2(b), the last two steps of the knowledge transformation is proposed but haven't been implemented fully.

b) *Unstable State*: Once there is a failure, the application suspends execution (by virtue of the transformed code) and moves to an unstable state. In this state, the injected code takes charge of the program and added components take charge of appropriate functionalities to heal the application back to a stable state.

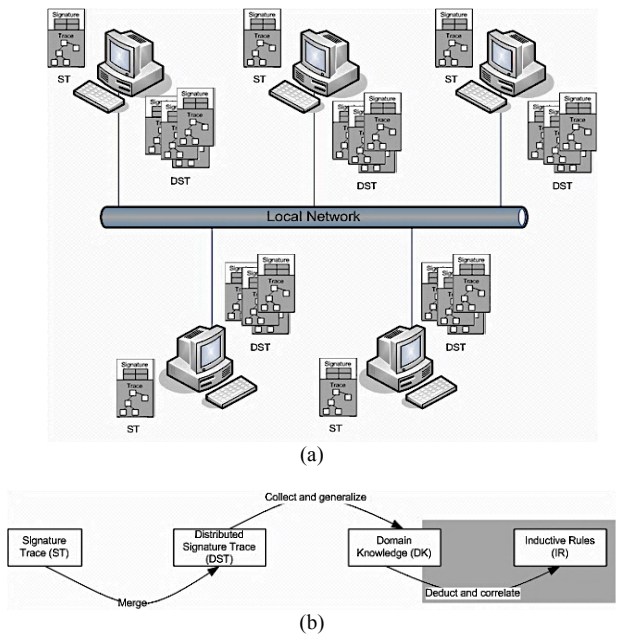

Fig. 2 System view and knowledge transformation.

In traditional self-management architecture, domain experts specify rules that map symptoms to fixes in an *if-then-else* format. Previously defined static rules might work well for simple systems where all possible failures are known in advance and a universal fix can quickly solve most failures. To extend such static approach, we proposed a new concurrent algorithm that employ feedback-driven loop to find the best solution scenario (s) for previously unseen failure. The overall algorithm to find fixes is presented bellow. Step 1 and 2 are concurrent processes for better performance of the algorithm. As noted earlier, after couple of runs, *DST* is replaced by *DK* and therefore the algorithm only mentions about *DST*.

1. Collect local application's signature trace (ST).
    A. Merge $ST_{i..n}$ to form Distributed Signature Trace (DST)
    B. In regular intervals, share $DST_i$ with other machines in the domain running the same software.
    C. If a new DST is received, then update local $DST_i$ with more generalized scenarios from the received DST.
2. If a failure, f is detected:
    A. Save the current system state and treat the system as unstable.
    B. The fault matching algorithm analyze the failure signature trace ($ST_c$) and match it with all possible DST and generate a list of fixes $f_1....f_n$ with existing success rate of those fixes.
    C. The list of possible fixes will be applied one ($f_x$) at a time to the application.
        i. If $f_x$ results in a stable state, increase the success rate of that ST and continue to step H.
        ii. If $f_x$ results in an unstable state and part of the DST exactly matches the $ST_c$, then apply the corresponding fix for that part of DST and continue as step C.i. The algorithm has to be aware of the depth of this kind of recursive try and should have a threshold value.
        iii. If $f_x$ results in an unstable state and there is a DST, which partially matches the $ST_c$, then mark that as a candidate ST for future processing.
    D. For all candidate ST, find the subset (size n) with the highest success rate.
    E. Calculate the distance of $ST_c$ with the members of the subset and find the ST with the lowest distance. Different distance formulas can be used to calculate this distance.
    F. Apply the newly found fix (in the closest ST) and
        i. if it results in a stable state, continue to step G, otherwise,
        ii. if not exceeded time limit (threshold) then refresh DST with others in the system and continue to step C
        iii. otherwise continue to step H.
    G. Resume the application.
    H. If all of the above fails then save the current status of the application, let the administrator know about the fault and log of the injected code.

Not all faults can be healed automatically or even recovered from. This paper is concerned with transient faults (network outage, memory overload, disk space outage etc.) that occur after the program is deployed. Such faults could result from problems in the user code, in the underlying physical system or network connection or in the run-time environment. Non-transient faults, caused by bugs in the user code (logical errors), user generated custom exceptions or faults generated due to the functional aspect of the program are outside the control of this approach and should be addressed by the system administrator or the developer of the user program.

## IV. FAULT SIMILARITY AND RELATIONSHIPS

Along with building *DST*s, the system also develops fault models to find similarity between a known fault and an unseen fault. A fault model is the collection of generalized signature traces and possible fixes for a specific fault scenario. Once a fault has occurred, it is analyzed and related *ST*s are tagged with that particular fault. For well-known faults, domain engineers can specify fault models early at the life cycle of the program. With more execution and fault occurrences, model for each fault will get enriched with the different execution scenarios and associated signature-traces. Once there are a substantial number of fault models acquired, a previously unseen fault can be matched with the faults in the model database by analyzing how similar each *ST* of two individual fault models is. As shown in Fig. 3 (a), the signature-traces of each fault in the fault model databases are matched and depending on that, the faults are arranged in a two-dimensional acyclic graph with weighed edges as shown in Fig. 3 (b). The measurements, which have to be made for the classification algorithm to work is as follows:

- *Positive match*: Number of *STs* that matched (within a certain error margin) between faults and if number of positive matches is greater than negative matches.
- *Negative match*: Number of incomparable STs and if number of negative matches ate higher than positive matches.
- *Cannot match*: Faults are from two different classes of objects with no common structure between them.
- *No match*: Faults from same class of objects without match in their *STs* or missing attributes in their signature.

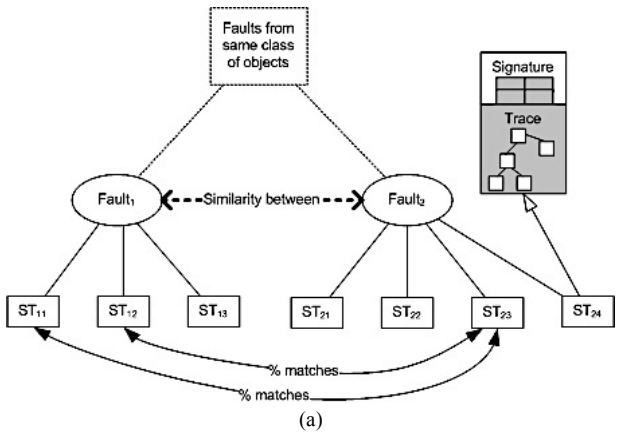

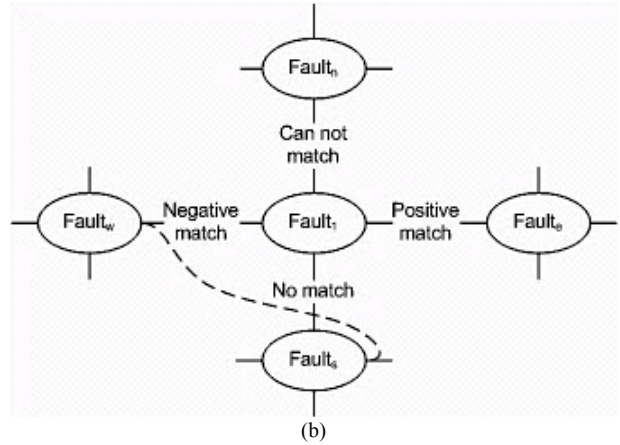

(b)

Fig. 3 Fault similarity and relationships.

Generally, faults from the same class of objects show positive matches, however, because of similar patterns in execution path and environment parameters, sometimes two different classes of faults can have a positive match between them (as shown by the dotted line). This complicates the graph and the matching process as this leads to a multiple dimension graph, intertwined with links between nodes in different dimensions. To minimize this situation, we added the following relationship refinement:

- Family tree: If the fault classes have semantic relationships, such as inheritance. This minimizes (can-not match and no-match cases) or strengthens (negative or positive match cases) links between faults. Faults from the same parent class are treated specially and commonalities are noticed between two sibling fault models for deducing a fix.

Once there is a stable fault model graph in the system, the following algorithm finds a match for an unknown fault.

```
// Argument: A fault, f, which is to be matched
// Returns: Fix, Fx
classify (fault f)
```
1. Put every fault model entity in a candidate set.
2. For every candidate, $F_c$ in the fault model, do
2.1. Calculate the percent match between $F_c$ and f.
   a) if exact match (within some error margin), returns find_fix($F_c$, f)
   b) if positive match, then discard all negative matched entities of $F_c$ from the set and record this percent match.
   c) if negative match, then discard all positive matched entities of $F_c$ from the set and record this percent match
   d) if can-not match or no-match then ignore.
3. Sort all percent match and
   a) pick the top most positive matched fault, $F_p$ and return find_fix($F_p$, f).
   b) if there are no positive matches, then select the lowest negative matched fault $F_n$ and return find_fix($F_n$, f).
   c) if no match at all, inform system administrator for intervention.

*find_fix(..)* matches the fault scenario with the candidate fault model and find the attached fix with that model. At the beginning, when the fault model is being developed, there will be a high degree of mismatches and frequent invocation to the system administrator. However, gradually this will improve as the fault model evolves to more matured and established state.

## V. EXPERIMENTATION AND EVALUATION

To evaluate the effectiveness of the transformed self-managed system, a robust and quantitative benchmarking methodology is needed. However, developing such a benchmark methodology is a non-trivial task [10] given the many evaluation issues and environment criteria to be resolved. If there were such a benchmarking methodologies, then the effectiveness will be quantitatively measured by comparing the performance of the proposed technique against benchmarks associated with the current modes of operation (without the self-healing approach). However, since that is not feasible with the current state of benchmarking technologies for self-managed systems; we can deploy techniques, such as [11], to measure the effectiveness of the proposed algorithm.

Java enterprise application server GlassFish [12] is selected as the target application and Java Platform Debugger Architecture [13] is being used to gather runtime application status to generate the program signatures and traces. Inside GlassFish application server, we run a custom server program within which we injected different faults to check our approach. This configuration was then cloned in multiple machines in the system and was execute simultaneously. We generate *ST*s from GlassFish and we crashed the custom server program inside GlassFish with injected faults so that we can match GlassFish's execution during a stable and unstable run. The custom program, which is run inside GlassFish, is a simple server that accepts incoming client request and serves those requests. By injecting faults in that server program, we can see what GlassFish does and develop and test our algorithm.

At first, we measure and analyze the time it takes for gathering and storing *ST*s and merging them into DST. Fig. 4 shows the timing requirements to gather ST and merge them to form DST. To calculate the times, we run five clone instances of the GlassFish in five different machines, ten times and then average was taken. The overhead to collect ST is negligible however the time requirement to merge STs into DST is substantial. To overcome this, we can stop tracing a given pathway any further, when there is a stable fault model in place that include this pathway. Also, instead of sharing the whole DST across the machine for updating purpose, an incremental updating process where only the latest additions will be distributed have to be devised. This will certainly minimize the overhead.

Next we looked into the affect of this overhead with relation to the size of the DST. As it is evident from Fig. 5, the overhead to gather ST is nearly constant regardless off overall size of the DST. However, as the size of the DST increases, the overhead to merge them also increases. As described above, an incremental merging algorithm will minimize this time for bigger DST. One thing to notice in this graph is that each size of the DST in the horizontal axis is gathered after different number of total runs of GlassFish (1, 10, 50, 100 and 500 total runs).

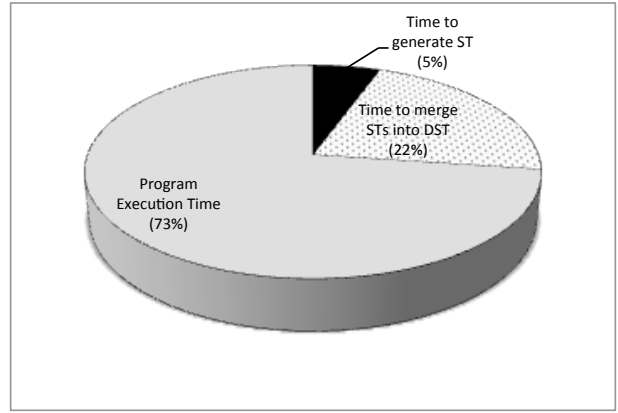

Fig. 4 Time requirements for generating ST and DST.

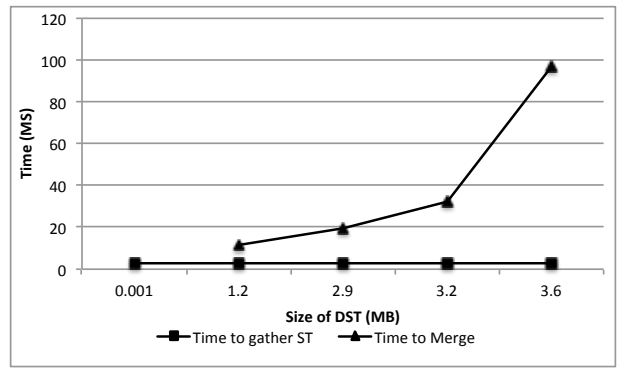

Fig. 5 Effect of DST size on timing requirement.

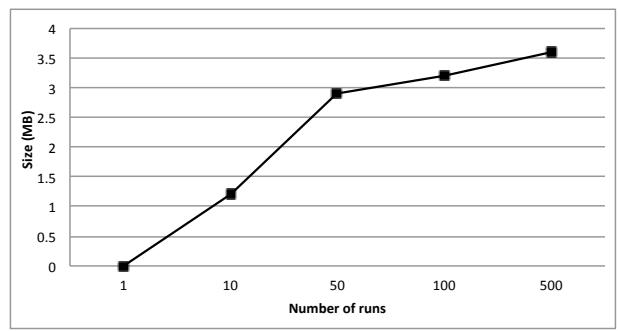

Fig. 6 Space requirement of DST.

Fig. 6 shows the space requirements for the resultant DST. Although with smaller number of runs, the space requirement is greater, it eases a lot as the number of runs increases and similar STs are combined together. We anticipate that the size of the DST will hover around a constant size after enough runs of the software. One thing to remember here is to reduce the communication overhead; incremental propagation of the DST is necessary across the machines instead of broadcasting the entire DST across the machines.

To find how efficiently the matching algorithm works, we injected a fault for which the ST is already established in the DST (manually inserted for testing). Since we know that the algorithm should return a match for that, we measure how long it took to do that on different size of DST. The matching time for a known fault was proportional to the size of the DST. With more occurrence of the same fault, the algorithm ranks the faults (and associated STs), so later occurrences can be matched much faster. To find an unknown fault we devised experimentations, which is yet to be implemented.

Several issues that need to be addressed for such experimentations, which we plan to investigate in the future, are:

- How to deviate a ST from a known ST and devise and inject corresponding fault?
- How to devise an accuracy matrix for partial match?
- How to find the timeout threshold to match unknown faults?
- Should timing requirement be addressed in case a unknown fault is categorized within a certain threshold?

Our goal is not to make the program run without any human intervention. We think that is impractical and even unattainable. We envision a system where the software heals autonomically most mundane and repeated faults themselves and only invoke the system administrator in case of completely new or miss-matched faults. That would certainly minimize system administrator's workload and will allow him/her to concentrate more high-level management issues. The technique presented in this paper will allow that.

## VI. CONCLUSIONS

Day-to-day maintenance of software systems is a grand challenge due to the fact that the runtime environment changes continuously. Users of such systems want to run their application and do not want to worry about the mundane task of system management in the face of a failure. If such management scenarios come into existence, the user wants the runtime environment to handle those situations autonomically. This paper presents a new technique of matching unknown fault scenarios to already established fault models to self-heal user applications. By capturing runtime parameters and execution pathways, stable execution models are established and later are used to match with an unstable execution scenario. All these support is provided transparently and the added functionalities are incorporated into existing user application by appropriate code transformation techniques. Results from experimentations are also presented that showed the usability of the proposed technique for self-healing applications.